\documentclass[12pt, fleqn]{article}
\usepackage[dvips]{graphicx}
\usepackage{amsmath,amssymb}
\author{Velin G. Ivanov{\footnote{Email address: vgosp@issp.bas.bg}} \\ \small{ \textsl{CP Laboratory, G. Nadjakov
Institute of Solid State Physics,} }
 \\ \small{ \textsl{Bulgarian Academy of Sciences, BG-1784 Sofia, Bulgaria} }}
\title{\textbf{Effects of constraints on the phase transition to
Bose-Einstein condensation}}
\date{}
\begin{document}
\parindent=0pt
\parskip=6pt
\rm \maketitle

\begin{abstract}
Classic and recent results for the critical behaviour of ideal
Bose gas at constant volume and constant pressure and for various
spatial dimensionalities $d>0$ are reviewed. New results about the
critical properties in a close vicinity of the $\lambda-$point are
presented.
\end{abstract}


\section{Introduction}
The thermodynamic potential and the correlation functions of ideal
Bose gas (IBG)~\cite{Landau:1980, Uzunov:1993} were studied in a
number of preceding papers~\cite{Gunton:1968, Cooper:1968,
Gunther:1974, Lacour:1974, Busiello:1985, Sun:1997, Simkin:1999,
Schakel:2003}; see also the review~\cite{Shopova:2003}. In this
paper we shall present a brief review of known results together
with an unified treatment of critical properties of IBG based on
the derivation of effective free energy of Landau-Ginzburg type.
Our consideration makes possible to outline a general picture of
the critical behaviour of IBG as well as to present some new
results about the effects of thermodynamic constraints: the
conditions of constant volume ($V = const$) and constant pressure
($P= const$) for certain spatial dimensionalities ($d >0$). We
should emphasize that the thermodynamic behaviour of IBG near the
$\lambda-$point corresponds either to a phase transition of first
order or, in cases which we shall enumerate, resembles a
continuous phase transition that is not equivalent to the standard
second order phase transition described by~$\varphi^4-$model (see,
e.g., Ref.~\cite{Uzunov:1993}). Our results could be used in the
interpretation of thermodynamic behaviour of extremely dilute
Bose-Einstein condensates (BEC), where the interparticle
interaction can be neglected.

\section{Thermodynamic potential}
We consider noninteracting and homogeneous gas of spinless bosons,
described by the Hamiltonian~\cite{Gunton:1968, Cooper:1968,
Lacour:1974}
\begin{equation}\label{1}
\hat{H}=\sum_{\vec{k}}\left[(\epsilon_k+r) \hat{a}_{\vec{k}}^{+}
\hat{a}_{\vec{k}}-\frac{1}{\sqrt{N}}
(\hat{a}_{0}^{+}h+\hat{a}_{0}h^{*})\right],
\end{equation}
where $r = -\mu \geq0$ is the modulus of the chemical potential
$\mu$, $\vec{k} = \{k_j; j = 1,...,d\}$ is the $d$-dimensional
wave vector, $\hat{a}_{\vec{k}}$ and $\hat{a}^{+}_{\vec{k}}$ are
second quantization operators, and $N$ is the total number of
bosons. The external field $h$ generates the spontaneous symmetry
breaking ($\langle \hat{a}_0\rangle
> 0$) below the critical temperature $T_c$. The energy spectrum
$\epsilon_k=ck^{\sigma}$ includes two cases: (i) $\sigma = 2$,
corresponding to real boson particles and boson excitations
produced by short-range interparticle (fermion and spin)
interactions, and (ii) $0 < \sigma < 2$, which stands for boson
excitations (composite bosons) produced by long-range
interparticle interactions (see, e.g., Refs.~\cite{Uzunov:1993,
Shopova:2003}). The parameter $c$ is equal to $(\hbar^2/2m)$,
where $m$ is the effective mass of the bosons.

The grand canonical thermodynamic potential of IBG,
\begin{equation}\label{2}
\Omega(T,r,h)=-\beta^{-1}\ln\mathrm{Tr}\mbox{ exp}\left( -\beta
\hat{H}\right),
\end{equation}
can be exactly calculated~\cite{Gunton:1968, Cooper:1968,
Lacour:1974}. This quantity can be written in general form

\begin{equation}\label{3}
\Omega(T,r,h)=-\beta^{-1}V\lambda_T^{-d}A(d,\sigma)g_{d/{\sigma}+1}(\beta
r )- \frac{1}{N}\frac{hh^{*}}{r},
\end{equation}
where $\lambda_T=(2\pi\hbar^2/mk_BT)^{1/\sigma}$ is the thermal
wavelength,

\begin{equation}\label{4}
g_{\nu}(y)=\frac{1}{\Gamma(\nu)}\int_0^{\infty}\frac{x^{\nu-1}}{e^{x+y}-1}dx
\end{equation}
is the integral Bose function (see, e.g., ~\cite{Uzunov:1993}) and

\begin{equation}\label{5}
A(d,\sigma)=\frac{2^{1-d+2d/{\sigma}}\Gamma(d/{\sigma})}{\sigma\pi^{d(1/2-1/{\sigma})}
\Gamma(d/2)}
\end{equation}
is a constant with the property $A(d,2)=1$. Note, that the grand
potential (3) obeys differential relation $d\Omega(T,r,h)=-SdT+ N
d{r}-\Psi dh^{*}-\Psi^{*} dh$, where $S$ is the entropy and
$(\Psi\sim\langle \hat{a}_0\rangle)$ is the order parameter of
BEC.

The free energy $\widetilde{\Omega}(T,r,\Psi)$, in which the
natural thermodynamic variable is the order parameter $\Psi$, can
be obtained from $\Omega(T,r,h)$ through the standard Legendre
transformation

\begin{equation}\label{6}
\widetilde{\Omega}(T,r,\Psi)={\Omega(T,r,h)+h\Psi^{*}+h^{*}\Psi
|}_{h(\Psi)},
\end{equation}
where

\begin{equation}\label{7}
\Psi=-\frac{\partial \Omega}{\partial
h^{*}}=\frac{h}{Nr},\hspace{0.3cm} \Psi^{*}=-\frac{\partial
\Omega}{\partial h}=\frac{h^{*}}{Nr}.
\end{equation}
Using Eqs.~(3)-(7) we obtain

\begin{equation}\label{8}
\widetilde{\Omega}(T,r,\Psi)=-\beta^{-1}V\lambda_T^{-d}A(d,\sigma)g_{d/{\sigma}+1}(\beta
r )+Nr{\Psi}^2.
\end{equation}

 We can restrict our calculations without a loss of generality to real values of $h$
and $\Psi$. The susceptibility is given by
$\chi_{T}=\partial\Psi/\partial h=1/Nr\sim t^{-\gamma}$, where
$\gamma$ is the susceptibility critical exponent and
$t=(T-T_c)/T_c$ (see, e.g., ~\cite{Uzunov:1993}).

For small $r$ and energies $(\epsilon_k\ll k_BT)$ the correlation
function $\chi(k)=\beta N^{-1}\langle
\hat{a}_{\overrightarrow{k}}^{+}
\hat{a}_{\overrightarrow{k}}\rangle$ takes the form
$N^{-1}(ck^{\sigma}+r)^{-1}$.
 This gives the correlation length $\xi=(c/r)^{1/\sigma}\sim t^{-\nu}$ and
$\chi(k)\sim k^{-\sigma}$ at $r=0$. From $\chi(k)\sim k^{-\sigma}$
we obtain that the Fisher exponent $\eta$, defined by $\chi(k)\sim
k^{-2+\eta}$, is equal to ($2-\sigma$) for all dimensional ranges
and constraints (see Table 1). In order to obtain the correlation
length exponent $\nu$ and the exponent $\gamma$ of susceptibility
$\chi_T$, we need to derive the function $r(t)$ for a general
spatial dimensionality $d$ for cases of constant volume and
constant pressure. The nullifying of chemical potential defines
the critical temperature $T_c$ of BEC for both cases ($V=const$,
$P=const$).

\section{Constant volume}
\begin{center}
\begin{figure}[t]
\includegraphics[angle=-90, width=0.7\textwidth]{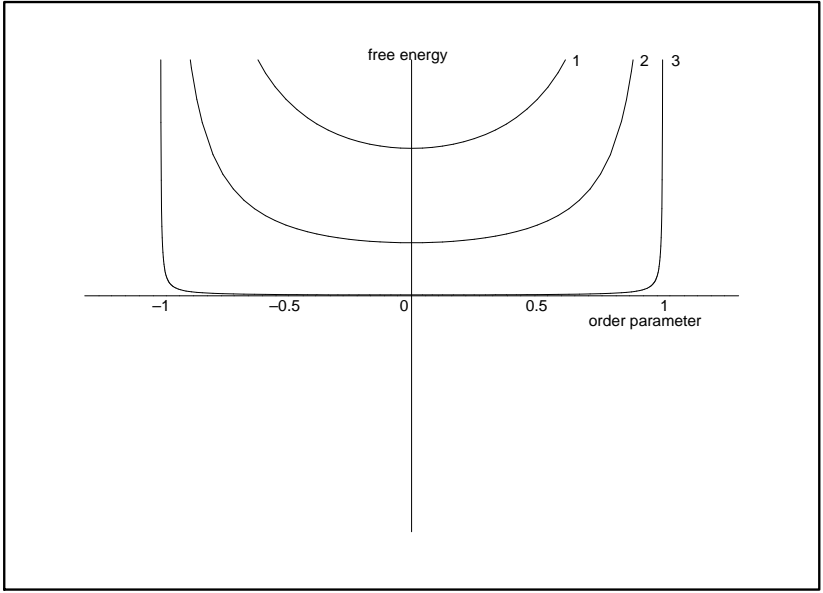}
\caption{\footnotesize Effective free energy for $d<\sigma$ $(d=1,
\sigma=2)$, $V=const$: line 1 ($T=0.25$), line 2 ($T=0.15$), line
3 ($T=0.02$). The case $d=\sigma$ is qualitatively the same.}
\end{figure}
\end{center}

\hspace{0.8cm}For the thermodynamic condition of constant volume
we need the free energy $f(T,N/V,\Psi)$, in which the natural
variable is the volume $V$ (or the number density $N/V$). To
obtain it we expand the integral Bose function $g_{\nu}(y)$ in
powers of $\beta r$ and express the chemical potential as a
function of the volume (or density). We expand the function
$g_{\nu}$ in the potential $\widetilde{\Omega}$ to the same order
in $\beta r$. This allows to make the next Legendre transformation

\begin{equation}\label{9}
f(T,N/V,\Psi)=\widetilde{\omega}(T,r,\Psi)-\frac{N}{V}r|_{r(N/V)},
\end{equation}
where

\begin{equation}\label{10}
\frac{N}{V}=\frac{\partial \widetilde{\omega}}{\partial
r}=\lambda_T^{-d}A(d,\sigma)g_{d/{\sigma}}(\beta r
)+\frac{N}{V}{\Psi}^2,
\end{equation}
$f=F/V$ and $\widetilde{\omega}=\widetilde{\Omega}/V$. Subtracting
the free energy of the disordered phase $f_{dis}$ we obtain the
following effective free energies corresponding to several
dimensionality ranges:

\begin{multline}\label{11}
f(T,N/V,\Psi)\sim
\frac{T^{\sigma/(\sigma-d)}}{(1-\Psi^2)^{d/(\sigma-d)}}\hspace{2.2cm}
d<\sigma\\
\sim
T^2\mbox{exp}\left(-\frac{1-\Psi^2}{T}\frac{N}{\lambda_0^{-d}A(d,\sigma)V}\right)\hspace{0.73cm}
d=\sigma\\  \sim
-\left(-\Psi^2-\frac{d}{\sigma}t\right)^{d/(d-\sigma)}\hspace{1.5cm}
\sigma<d<2\sigma \\
\sim (\Psi^2+2t)^2\frac{1}{\ln(\Psi^2+2t)^{-1}}\hspace{1.9cm}
d=2\sigma\\ \hspace{-2cm} \sim
\left(\frac{\sigma}{d}\Psi^2+t\right)^2\hspace{4cm} d>2\sigma.
\hspace{5.3cm}
\end{multline}

 The case $d<\sigma$ is depicted in Fig.~1 for $d=1$ and $\sigma=2$. We take the
volume per particle $v=V/N$ to be unity and $m=\hbar=k_B=1$. The
number coefficient in Eq.~(11) is equal to $1/2$ and $T_c=0$. When
the temperature tends to zero the curve $f(\Psi)$ acquires a flat
part, indicating that the order parameter $\Psi$ occurs with a
jump at $T_c=0$. In Fig.~2 we show the $d>2\sigma$ case, when
$d=5$ and $\sigma=2$. Now the numerical factor in Eq.~(11) is 8.96
and the normalization $\widetilde{\Psi}=\sqrt{\sigma/d}\Psi$ has
been used. The function $f(\Psi)$, shown in Figs.~1 and 2, is
quite different from the respective potential $f(\Psi)$ known for
the usual cases of IBG ($d=3$, $\sigma=2$)~\cite{Cooper:1968} and
the~$\varphi^4-$model~\cite{Uzunov:1993}.

Using the free energy (11) we calculate the specific heat at
constant volume $C_V=-T(\partial^2 f/\partial T^2$) and the
equation of state $h=\partial f/\partial \Psi$. Bearing in mind
the definitions (see, e.g., ~\cite{Uzunov:1993}), $C_V\sim
t^{-\alpha}$, $\Psi\sim h^{1/\delta}$ $(h\sim 0)$ and $\Psi\sim
t^{\beta}$ $(h=0)$ we obtain the critical exponents $\alpha$,
$\delta$ and $\beta$. The values of the critical exponents are
given in Table 1{\footnote{In Table 1 we distinguish logarithmic
factors by "exponents" with suffix $l$. Thus $f(t)$ is said to
have exponent $x_l$ or $x^l$, respectively, if $f(t)\sim t^{x}/\ln
t^{-1}$ or $f(t) \sim t^{x}\ln t^{-1}$~\cite{Gunton:1968}.}}. The
behaviour of $\Psi$, $V$, $S$, $C_V$ and $\chi_T$ at $T_c$ is
given in Table 2.

\begin{center}
\begin{figure}[t]
\includegraphics[angle=-90, width=0.7\textwidth]{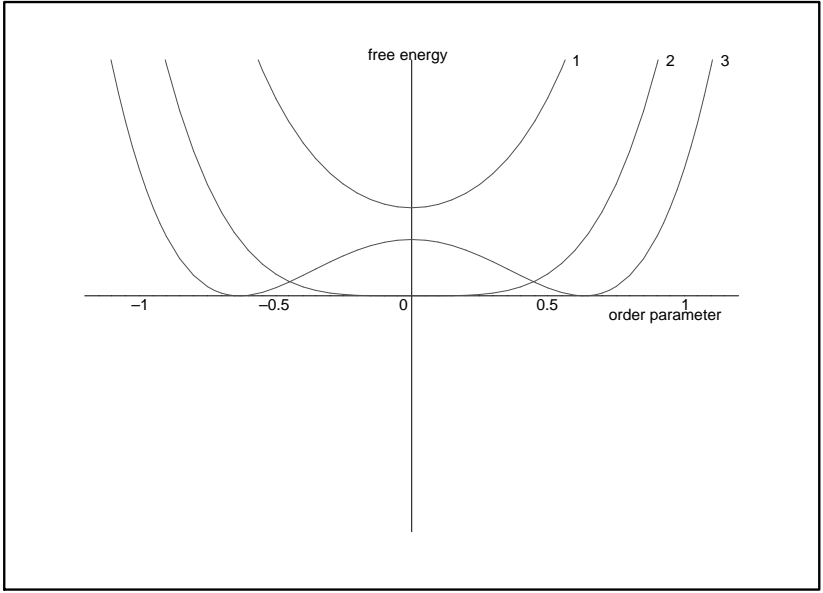}
\caption{\footnotesize Effective free energy for $d>2\sigma$
$(d=5, \sigma=2)$, $V=const$: line 1 ($t=0.5$), line 2 ($t=0$),
line 3 ($t=-0.4$). The case $d=2\sigma$ is qualitatively the
same.}
\end{figure}
\end{center}

\section{Constant pressure}
When the system is under constant pressure we should use the free
energy per particle $\mu$, in which the natural variable is the
pressure $P$ . To obtain the chemical potential we solve the
equation for the pressure $P=-\partial\widetilde{f}/\partial v$
with respect to the volume per particle $v$ to the lowest order in
$t$. Thus we make the following Legendre transformation

\begin{equation}\label{12}
\mu(T,P,\Psi)=\widetilde{f}(T,N/V,\Psi)+Pv|_{v(P)},
\end{equation}
where $\mu=\Phi/N$ and $\widetilde{f}=F/N=fv$.

For the various dimensional ranges we obtain

\begin{multline}\label{13}
\mu(T,P,\Psi)\sim
-(1-\Psi^2)t^{\sigma/d}(1+at)\hspace{2.9cm} d<\sigma\\
\sim-(1-\Psi^2)\frac{t+at^2}{\ln t^{-1}}\hspace{4.8cm} d=\sigma\\
\sim-(1-\Psi^2)t(1+at^{d/\sigma-1})\hspace{2.65cm} \sigma<d<2\sigma\\
\sim-(1-\Psi^2)t+(1-\Psi^2)t^2(1+at)\ln t^{-1}\hspace{0.95cm}
d=2\sigma\\ \hspace{-3cm}
\sim-(1-\Psi^2)t-(1-\Psi^2)t^2(1+at^{d/\sigma-2}+bt)\hspace{0.3cm}
d>2\sigma.\hspace{3.7cm}
\end{multline}

In Fig.~3 we depict the case $\sigma<d<2\sigma$ for $d=3$ and
$\sigma=2$. We set again $m=\hbar=k_B=1$ and take the magnitude of
the pressure $P$ so as to give $T_c=1$. This gives the factor
$1.28$ for the coefficient of proportionality in Eq.~(13). This
picture should be compared with those shown in Figs.~1 and 2 (see,
also, ~\cite{Uzunov:1993, Cooper:1968}).

\hspace{0.4cm}We derive from the free energy $\mu$ the specific
heat at constant pressure $C_P=-T(\partial^2 \mu/\partial T^2$)
and the exponent $\alpha$. Moreover the chemical potential (13)
can be used to calculate the correlation length and susceptibility
exponents $\nu$ and $\gamma$. Since the order parameter has a
jump, as is seen from Eq.~(13) ($\mu$ is zero below $T_c$), the
exponents $\beta$ and $\delta$ cannot be defined. The values of
the critical exponents and the behaviour of $\Psi$, $V$, $S$,
$C_P$ and $\chi_T$ at $T_c$ are given in Tables 1 and 2.

\begin{center}
\begin{figure}[t]
\includegraphics[angle=-90, width=0.7\textwidth]{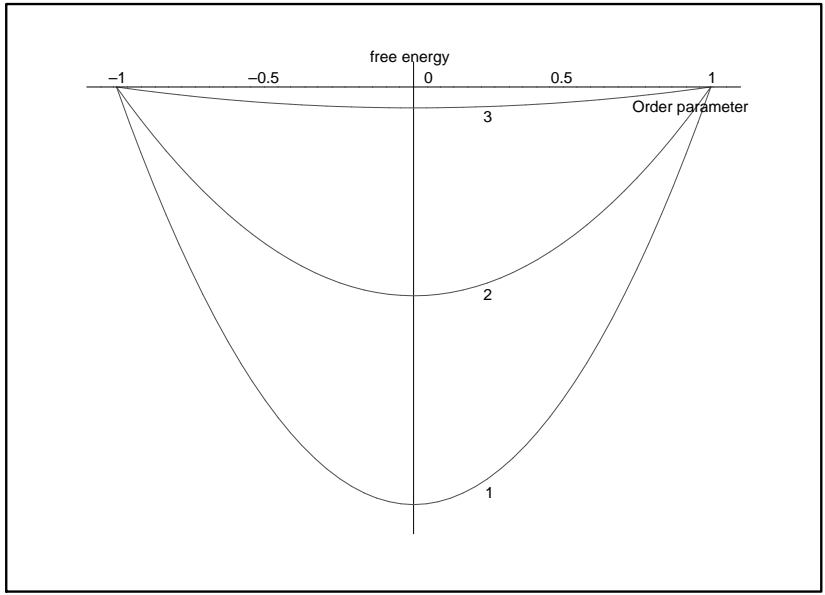}
\caption{\footnotesize Effective thermodynamic potential for
$d>\sigma$ $(d=3, \sigma=2)$, P=const: line 1 ($t=0.1$), line 2
($t=0.05$), line 3 ($t=0.005$). The case $d\leq\sigma$ is
qualitatively the same.}
\end{figure}
\end{center}

\section{Summary}
In this paper we derived and analyzed the IBG free energies of
Landau-Ginzburg type for the thermodynamic constraints of constant
volume and constant pressure at different spatial dimensionalities
$d>0$. The values of the critical exponents  for various spatial
dimensionalities are given in Table 1. The critical exponents for
$d=(\sigma,2\sigma)$ and $P=const$ are new results of our
investigation. The derived free energies enable us to present the
behaviour of thermodynamic functions at the critical point. It is
shown in Table 2. Besides, our results make possible to estimate
the order of the phase transition to BEC. When $d>\sigma$ and
$P=const$ this phase transition is of first order. But for lower
dimensionalities ($d\leq\sigma$) and in both cases ($P=const,
V=const$), the phase transition to BEC resembles a first order
phase transition only by the order parameter jump. On the other
hand, for $d>\sigma$ and $V=const$ the phase transition to BEC is
a continuous phase transition with properties, which are quite
different from the properties of the standard second order phase
transition described by~$\varphi^4-$model. Our results can be used
in the interpretation of experimental results for BEC in extremely
dilute boson gases at constant pressure or volume.

\begin{center}
\small{} \textsc{Table 1}. \textmd{Critical exponents of the ideal
Bose gas.}

\begin{tabular}[h]{|r|c|c|c|c|c|c|l|}
      \hline $\textsc{volume=}\textmd{const}$  & $\alpha$ &  $\alpha_s$ & $\beta$ & $\gamma$ & $\delta$ & $\nu$ & $\eta$ \\
      \hline $d<\sigma \quad (T_c=0)$ & $-d/\sigma$ &  $-$ & $*$ & $\frac{\sigma}{\sigma-d}$ & $*$ & $\frac{1}{\sigma-d}$ & $2-\sigma$ \\
      \hline $d=\sigma \quad (T_c=0)$ & $-1$ &  $-$ & $*$ & $\infty$ & $*$ & $\infty$ & $2-\sigma$ \\
      \hline $\sigma<d<2\sigma \quad (T_c>0)$ & $0$ &  $\frac{d-2\sigma}{d-\sigma}$ & $1/2$ & $\frac{\sigma}{d-\sigma}$ & $\frac{d+\sigma}{d-\sigma}$ & $\frac{1}{d-\sigma}$ & $2-\sigma$ \\
      \hline $d=2\sigma \quad (T_c>0)$ & $0$ &  $0_{l}$ & $1/2$ & $1_{l}$ & $3_{l}$ & $\sigma\nu=1_{l}$ & $2-\sigma$ \\
      \hline $d>2\sigma \quad (T_c>0)$ & 0 &  $(\frac{2\sigma-d}{\sigma}, -1)^a$ & $1/2$ & 1 & 3 & $1/\sigma$ & $2-\sigma$ \\
      \hline
      \hline $\textsc{pressure=}\textmd{const}$  & $\alpha$ &  $\alpha_s$ & $\beta$ & $\gamma$ & $\delta$ & $\nu$ & $\eta$ \\
      \hline $d<\sigma \quad (T_c>0)$ & $2-\sigma/d$ &  $-1^b$ & $*$ & ${\sigma}/{d}$ & $*$ & ${1}/{d}$ & $2-\sigma$ \\
      \hline $d=\sigma \quad (T_c>0)$ & $0_{l}$ &  $-$ & $*$ & $1_{l}$ & $*$ & $\sigma\nu=1_{l}$ & $2-\sigma$ \\
      \hline $\sigma<d<2\sigma \quad (T_c>0)$ & $2-d/\sigma$ &  $-$ & $*$ & $1$ & $*$ & $1/\sigma$ & $2-\sigma$ \\
      \hline $d=2\sigma \quad (T_c>0)$ & $0$ &  $0^l$ & $*$ & $1$ & $*$ & $1/\sigma$ & $2-\sigma$ \\
      \hline $d>2\sigma \quad (T_c>0)$ & 0 &  $(\frac{2\sigma-d}{\sigma}, -1)^a$ & $*$ & 1 & $*$ & $1/\sigma$ & $2-\sigma$ \\
      \hline
\end{tabular}
\end{center}
\footnotesize{*The order parameter has a jump (from 0 to 1) at
$T_c$.}\\ \footnotesize{$^a$The first entry is for
$d<3\sigma$, the second is for $d>3\sigma$. }\\
\footnotesize{$^b$This entry is only for $d=\sigma/2$. }

\newpage
\begin{center}
\small{} \textsc{Table 2}. \textmd{Behaviour of $\Psi$, $V$, $S$,
$C_V$, $C_P$, $\chi_T$ at $T_c$.}

\begin{tabular}[h]{|r|c|c|c|c|l|}
      \hline $\textsc{volume=}\textmd{const}$  & order par. &  volume & entropy & heat cap. & suscept. \\
      \hline $d<\sigma \quad (T_c=0)$ & jump & const. & cont. & cont. & diverg. \\
      \hline $d=\sigma \quad (T_c=0)$ & jump & const. & cont. & cont. & diverg. \\
      \hline $\sigma<d<2\sigma \quad (T_c>0)$ & cont. & const. & cusp & cusp & diverg. \\
      \hline $d=2\sigma \quad (T_c>0)$ & cont. & const. & cusp & cusp & diverg. \\
      \hline $d>2\sigma \quad (T_c>0)$ & cont. & const. & cusp & cusp & diverg. \\
      \hline
      \hline $\textsc{pressure=}\textmd{const}$  & order par. &  volume & entropy & heat cap. & suscept. \\
      \hline $d<\sigma \quad (T_c>0)$ & jump & cont. & cont. & cont/jump/div.$^a$ & diverg. \\
      \hline $d=\sigma \quad (T_c>0)$ & jump & cont. & cont. & cont. & diverg. \\
      \hline $\sigma<d<2\sigma \quad (T_c>0)$ & jump & jump & jump & divergent & diverg. \\
      \hline $d=2\sigma \quad (T_c>0)$ & jump & jump & jump & ln divergent & diverg. \\
      \hline $d>2\sigma \quad (T_c>0)$ & jump & jump & jump & jump & diverg. \\
      \hline
\end{tabular}
\end{center}
\footnotesize{$^a$Respectively for
$d<\sigma/2$,\hspace{0.2cm}$d=\sigma/2$,\hspace{0.2cm}$d>\sigma/2$.
} \normalsize{}

\section*{Acknowledgments}
The author thanks Prof. Dimo I. Uzunov for valuable discussions
and critical reading of the manuscript.

\end{document}